\documentclass{aa}  

\usepackage[switch]{lineno}
\usepackage{graphicx}
\usepackage{comment}
\usepackage{txfonts}
\usepackage{color}
\usepackage{comment}
\usepackage{soul}
\begin{document} 

   \title{Evidence for protons accelerated and escaped from the Puppis A region using \textit{Fermi}-LAT observations}

   \author{R. Giuffrida \inst{1,2,3}, M. Lemoine-Gourmard \inst{4}, M. Miceli \inst{1,2}, S. Gabici \inst{5}, H. Sano \inst{6,7}, M. Aruga \inst{6}, M. Mayer \inst{8}, W. Becker \inst{9,10}, M. Sasaki\inst{8},  Y. Fukui \inst{6}}

    \institute{Dipartimento di Fisica e Chimica E. Segrè, Università degli Studi di Palermo, Piazza del Parlamento 1, 90134, Palermo, Italy
         \and
             INAF-Osservatorio Astronomico di Palermo, Piazza del Parlamento 1, 90134, Palermo, Italy
        \and
            AIM, CEA, CNRS, Université Paris-Saclay, Université de Paris, F-91191 Gif sur Yvette, France
        \and
            Universitè Bordeaux, CNRS, LP2I Bordeaux, UMR 5797, F-33170 Gradignan, France
        \and
            Université de Paris, CNRS, Astroparticule et Cosmologie, F-75013 Paris, France
        \and 
            Department of Physics, Nagoya University, Furo-cho, Chikusa-ku, Nagoya, Aichi 464-8601, Japan
        \and
            Faculty of Engineering, Gifu University, 1-1 Yanagido, Gifu, Gifu 501-1193, Japan
        \and
            Dr. Karl Remeis-Sternwarte and Erlangen Centre for Astroparticle Physics, Friedrich-Alexander Universit$\Ddot{\rm{a}}$t Erlangen-N$\Ddot{\rm{u}}$rnberg, Sternwartstraße 7, 96049 Bamberg, Germany
        \and
            Max-Planck-Institut für extraterrestrische Physik, Giessenbachstrasse, 85748 Garching, Germany
        \and
            Max-Planck Institut für Radioastronomie, Auf dem Hügel 69, 53121 Bonn, Germany
    }

% \abstract{}{}{}{}{} 
%context

% 5 {} token are mandatory
 
\abstract
{{Supernova remnants (SNRs) interacting with molecular clouds are interesting laboratories to study the acceleration of cosmic rays and their propagation in the dense ambient medium. We analyze 14 years of Fermi-LAT observations of the supernova remnant Puppis A to investigate its asymmetric $\gamma$-ray morphology and spectral properties. This middle-aged remnant ($\sim$4 kyr) is evolving in an inhomogeneous environment, interacting with a dense molecular cloud in the northeast and a lower-density medium in the southwest. We find clear differences in both $\gamma$-ray luminosity and spectral energy distribution between these two regions. The emission from both sides is consistent with a hadronic origin. However, while the southwestern emission can be explained by standard Diffusive Shock Acceleration (DSA), the northeastern side may involve re-acceleration of pre-existing cosmic rays or acceleration via reflected shocks in the dense cloud environment. Additionally, we identify two significant $\gamma$-ray excesses outside the remnant, including a previously unreported source to the south. These features are likely produced by cosmic rays that have escaped Puppis A and are interacting with nearby dense molecular material. From this extended emission, we estimate the total energy in escaping cosmic rays to be $W_{CR} \sim 1.5 \times 10^{49}$ erg, providing important constraints on cosmic-ray propagation around the remnant.}}

   
   \titlerunning{Cosmic ray acceleration in Puppis A}
   \authorrunning{Giuffrida et al.}

   \maketitle

\section{Introduction}
\label{sec:intro}
The $\gamma-$ray emission of supernova remnants (SNRs) provides a direct evidence of particle acceleration at their shock fronts. Two different physical processes can be invoked for the production of $\gamma-$ray emission, namely the leptonic and hadronic scenarios. The leptonic case involves emission from ultrarelativistic electrons via inverse Compton (IC) and/or bremsstrahlung. On the other hand, in the hadronic scenario, proton acceleration can be studied via pion decay associated with the impacts of high-energy hadrons with the ambient medium: proton-proton interaction produces a neutral pion which subsequently decays in two gamma rays. 
This emission can be enhanced in SNRs evolving in a dense interstellar medium, where the accelerated particles interact with the dense gas. 
This makes SNRs interacting with molecular clouds (MCs) interesting laboratories to study hadron acceleration. 
Another mechanism which can be responsible for $\gamma-$ray emission from SNRs interacting with dense MCs takes into account the reacceleration process. In particular, \cite{ubf10} have shown that the $\gamma$-ray emission from middle-aged SNRs can be explained with the crushed cloud scenario, in which the $\gamma$-ray emission is due to radiatively-compressed shocked clouds.  Also bright synchrotron radio emission is expected to stem from this “crushed cloud" \citep{bc82}. Examples are W51C, W44 and IC 443 \citep{ubf10}.
In particular, the multi-wavelength spectrum of W44 can be well explained by reaccelerated particles alone, without requiring any spectral break—only a high-energy cutoff corresponding to the maximum energy achievable by the accelerator \citep{2016A&A...595A..58C}.
Moreover, particles escaping from the SNR shock might be responsible for $\gamma$-ray emission in nearby interstellar clouds, thus providing an indirect evidence of cosmic-rays acceleration and escape from the SNR \citep{aa96,ga07,gac09,omy11,mrc21}.

\begin{figure*}[ht!]
    {\includegraphics[width=\textwidth]{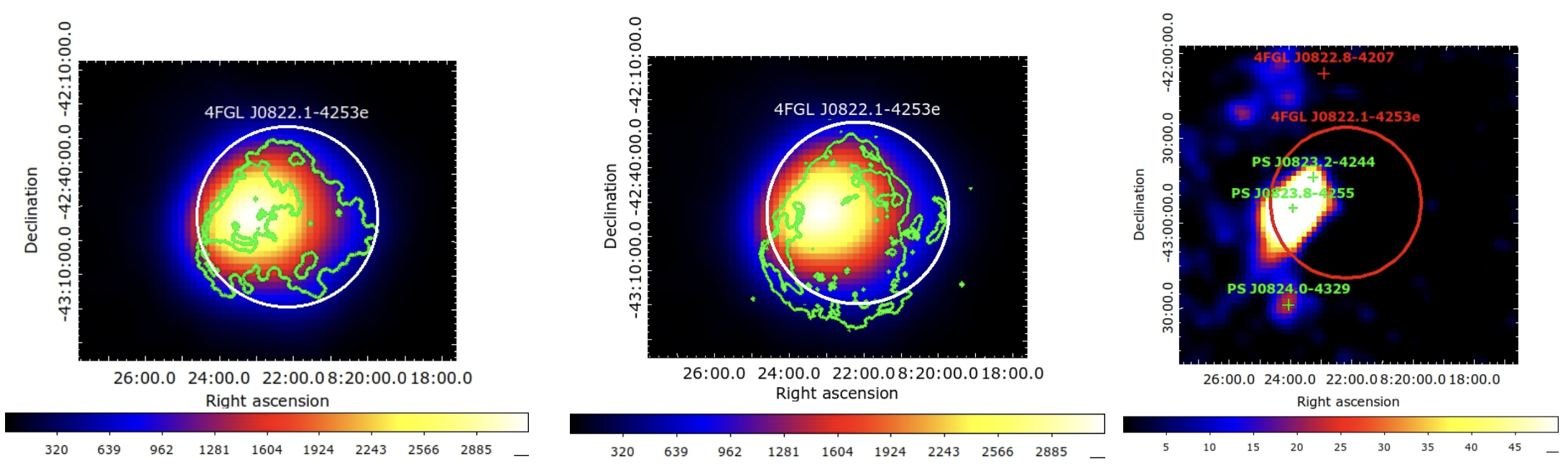}}
    \caption{\textit{Left panel:} Fermi-LAT $TS$ map showing the $\gamma$-ray emission of Puppis A. The white circle marks the uniform disk used to fit Puppis A in 4FGL-DR3. The green contours mark the contour levels at the 1\% and 10\% of the maximum of the \textit{eROSITA} map \citep{mbp22}. \textit{Central pane}: same as left panel but with radio VLA contours at 7\% and 15\% of the maximum of the VLA map. \textit{Right panel: } Residual TS map obtained by fitting the ROI (see text) using the disk template shown in the upper panel. The red circle shows the uniform disk which models Puppis A (same as white circle in the upper panel), the red cross shows the source 4FGL-J0822.8-4207 and the green crosses indicate the new sources needed to fit the gamma ray excesses (Table \ref{tab:PS}).}
    \label{fig:Resid_disk}
\end{figure*}

Puppis A is a 4 kyr old \citep{1988srim.conf...65W,bpw12,2020ApJ...899..138M}, shell-like Galactic SNR, at a distance of about 1.3 kpc \citep{rcw17}, with an angular diameter of approximately 50' (around 11 pc). It has been observed at different energy bands such as radio, X-rays and $\gamma$-rays. The radio emission of Puppis A has been analyzed using \textit{VLA} observations \citep{dbw91,cdg06} and is characterized by very bright synchrotron radiation in the eastern side of the remnant (see green contours in the central panel of Fig. \ref{fig:Resid_disk}). The eastern region is also characterized by a high surface brightness in X-rays and includes the Bright Eastern Knot (BEK), which is indicative of shock-cloud interaction \citep{hfp05}. The asymmetry in the X-ray emission has been confirmed by \cite{dlr13} and \cite{mbp22}, revealing that the remnant is evolving in an inhomogeneous ambient medium. The X-ray emission of Puppis A is mainly dominated by thermal emission from shock-heated interstellar medium \citep{hfp05}, but also some isolated ejecta knots have been detected \citep{hpf08,kmt08,kwp10,mbp22}. 

Because of the interaction with interstellar clouds (see \citealt{da88,asf22}), Puppis A is an interesting source to study particle acceleration. 
In particular, the interaction with this complex ambient medium can also result in various parts of the remnant being in different evolutionary phases (as, for example, for the Cygnus Loop, \citealt{tba21}).
Previous works have studied the $\gamma$-ray emission of Puppis A with the \textit{Fermi}-LAT telescope, using 4 years of observations for \cite{hgl2012} and 7 years for \cite{xgl17}, finding an asymmetry in the morphology of the source (the eastern side being brighter than the western side) that is consistent with the X-ray morphology. The $\gamma$-ray spectrum analyzed in \cite{hgl2012} can be fitted with a power law with an index $\Gamma=2.1$ in the 200 MeV - 0.1 TeV energy range and some indications for a slightly softer emission in the western part of the remnant have been found, though the poor statistics available did not allow them to confirm it.
\cite{hgl2012} and \cite{xgl17} also estimated the total energy of accelerated particles of $E\sim5 \times 10^{49}$ erg and $7.5 \times 10^{49}$ ($n/4.0$ cm$^{-3}$) $^{-1}$ erg, respectively. In both previous works, a leptonic scenario for the $\gamma$-ray emission of Puppis A was disfavored because of the following reasons. An inverse Compton (IC)-dominated model demands both an unusually low ambient density and a significantly high electron-to-proton ratio. Similarly, a bremsstrahlung-dominated emission also requires an exceptionally high electron-to-proton ratio. Moreover, the spectral break observed at radio frequencies across the entire remnant implies a corresponding break in the electron spectrum, which is not reflected at $\gamma$-ray energies. Consequently, the authors concluded that the hadronic scenario is the most plausible explanation.

In order to study the origin of the $\gamma$-ray emission, \cite{HessColl15} explored the emission of Puppis A at very high energies (VHE), but without obtaining a significant detection. The non-detection of Puppis A by H.E.S.S. can be interpreted as a signature of shock velocity which is not large enough to accelerate particles up to TeV. 

The $\gamma$-ray emission of Puppis A has also been studied in \cite{agk2022}, using 14 years of \textit{Fermi}-LAT observations. They aim at studying the origin of the point-like $\gamma$-ray source 4FGL J0822.8-4207, suggesting that it can be associated either with cosmic-ray accelerated and escaped from Puppis A interacting with denser ambient medium or with the protostellar jet HH219.

In this paper, we report on the analysis of 14 years of \emph{Fermi}-LAT observations, which significantly improve the statistics with respect to the data analyzed by \citet{hgl2012,xgl17}. 
The paper is organized as follows: Sect. \ref{sec:data} describes the \textit{Fermi}-LAT observations and the data analysis; results on both the morphology of the source and the spectral analysis are shown in Sect \ref{sec:results}; discussions appear in Sect. \ref{sec:discussion} and finally the conclusions are presented in Sect. \ref{sec:conc}. 

\section{Observations and Data Analysis}
\label{sec:data}
We analyzed 14 years of observations carried out by \textit{Fermi}-LAT between August 2008 and September 2022 focusing on the region of interest (ROI) of $15^\circ \times 15 ^\circ$, centered on  4FGL J0822.1-4253e (Puppis A, RA = $8^h 24^m 6.96^s$, DEC = $-42^\circ 59' 49.2''$), with spatial bins of $0.03^\circ$ and 10 energy bins, including all sources from the Fermi-LAT 14-year Source Catalog (4FGL-DR4, \citealt{DR32022,DR4}) in a region of $25^\circ \times 25^\circ$. \texttt{PASS 8} data\footnote{see \url{https://fermi.gsfc.nasa.gov/ssc/data/analysis/documentation/Pass8_edisp_usage.html} for details} have been analyzed by using the software \texttt{fermitools\footnote{\url{https://fermi.gsfc.nasa.gov/ssc/data/analysis/software/}}} version 2.2.0 and the \texttt{Python} package \texttt{fermipy\footnote{\url{https://fermipy.readthedocs.io/en/latest/}}} \citep{fermipy} version 1.2 and filtered with \texttt{DATA\_QUAL>0, LAT\_CONFIG==1}. The background has been modelled taking into account the sources in the 4FGL-DR4 catalog, the Galactic diffuse emission (provided by the file \texttt{gll\_iem\_v07}) and the `Source' events of the \texttt{P8R3} instrument response functions (IRFs). 
We analyzed data between 1 GeV - 1 TeV for the morphological analysis (Sect. \ref{sec:morph}) and between 300 MeV - 1 TeV for the spectral analysis (Sect. \ref{sec:spec_puppisA}).  Only PSF3 events have been kept between 300 MeV and 1 GeV because they provide the best angular resolution to resolve the source, and all PSF event types were used above 1 GeV.
To reduce Earth's limb contamination, we selected only events with a zenith angle $< 90$° between 300 MeV and 1 GeV and $<105$° above 1 GeV. 

\section{Results}
\label{sec:results}
\subsection{Morphological Analysis}
\label{sec:morph}
In order to study the morphology of an extended source like Puppis A, we take advantage of the significant improvement of the \emph{Fermi-}LAT angular resolution at high energies, by analyzing data in 1-1000 GeV energy band only. 
The analysis has been performed by following the maximum likelihood method described in \cite{mbc1996}, which also provides the significance of each source as the square root of the Test Statistic ($TS$), which is given by $TS = 2 \log{\frac{L}{L_0}}$, where the likelihood $L$ is obtained by fitting the source model plus the background components (including other sources) to the data, while the likelihood $L_0$ is derived by fitting the background component only.
The analysis has been performed by freeing the spectral parameters of the galactic diffuse background, of the isotropic background and of all the sources located within 10° from the ROI center with $TS>100$.
Then, we adopted the tool \texttt{find\_sources} to localize new point-like sources with a $TS$ value above 16. 
The TS map of Puppis A is shown in the left panel of Fig. \ref{fig:Resid_disk}.

The right panel of Fig. \ref{fig:Resid_disk} shows the difference between the morphology of Puppis A and the uniform disk of the previous 4FGL-DR3 catalog. It can be noticed that using the model in the previous catalog 4FGL-DR3 \citep{DR32022}, where Puppis A is fitted as a uniform disk, a bright emission in the eastern part of the remnant is visible. This excess can be modelled by adding the ad hoc point-like sources found in \cite{agk2022}: PS J0823.2-4244 and PS J0823.8-4255. The presence of this excess in the eastern part of Puppis A (fainter than Puppis A but still significant at 5 $\sigma$ confidence level) shows that the uniform disk model does not describe the morphology of the source properly, because of an asymmetry in the emission of Puppis A. However, in the new catalog 4FGL-DR4 the morphology of Puppis A has been improved using the X-ray emission above 1 keV detected with \textit{eRosita} \citep{mbp22}. 
Moreover, we also found a $\gamma-$ray emission excess beyond the southern border of the remnant (visible in the right panel of Fig. \ref{fig:Resid_disk}), which can be well fitted by a point-like source (PS J0824.0-4329) with a $TS$ value of 22. The coordinates of these additional sources are summarized in Table \ref{tab:PS}.

\begin{table}[h!]
	\caption{Coordinates of additional pointlike sources for the uniform disk model.}
	\begin{center}
		\footnotesize
		\begin{tabular}{c c c}
		    \hline\hline
			Name & RA & DEC \\
			\hline
		    PS J0823.2-4244 & 8$^h$ 23$^m$ 13.04$^s$ & -42$^\circ$ 44$'$ 30.40$''$\\
            PS J0823.8-4255 & 8$^h$ 23$^m$ 52.21$^s$ & -42$^\circ$ 55$'$ 19.14$''$ \\
            PS J0824.0-4329 & 8$^h$ 24$^m$ 2.00$^s$ & -43$^\circ$ 29$'$ 31.20$''$ \\
            \hline
		\end{tabular}
		\label{tab:PS}
	\end{center}
\end{table}

\begin{figure*}[htb!]
\centering
    \includegraphics[width=\textwidth]{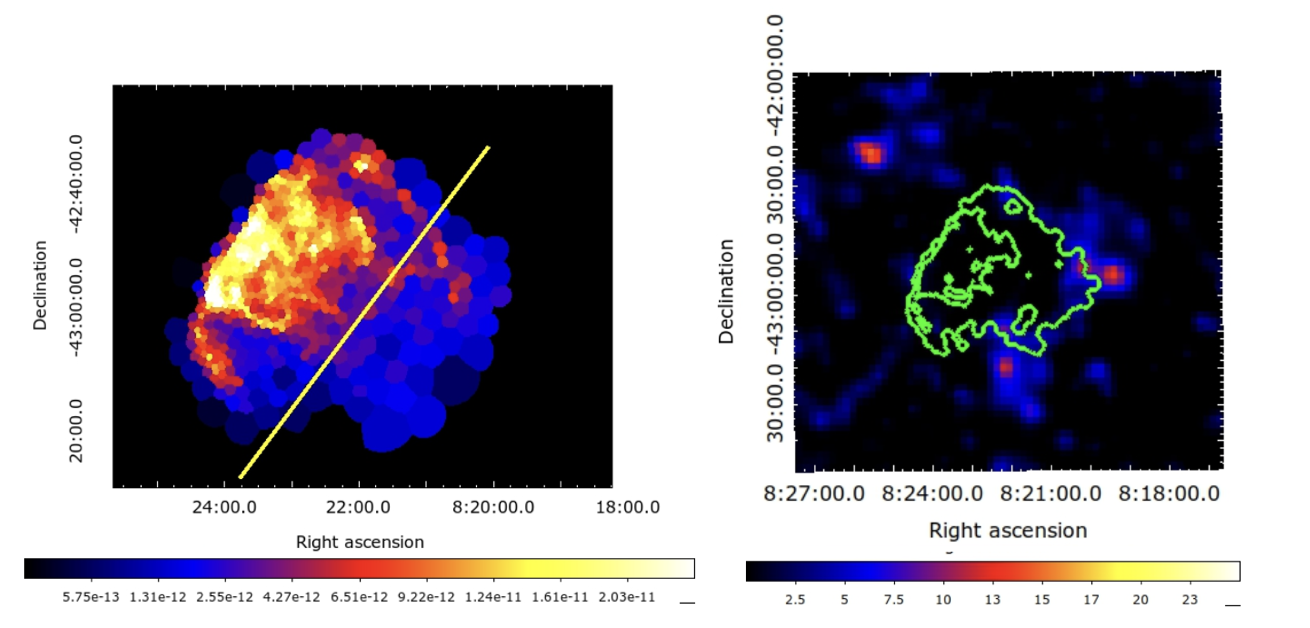}
    \caption{\emph{Left panel}: \textit{eROSITA} un-absorbed flux image in erg/s/cm$^2$/arcmin$^2$ in the $0.7-1$ keV band in square root scale. The yellow line marks the separation between northeastern and southwestern side (see text).  \emph{Right panel}: residual TS map where the morphology of Puppis A is fitted with the template in the upper panel and the source 4FGL J0822.8-4207 is considered as an extended source.}
    \label{fig:template_correct}
\end{figure*}

\begin{figure}[htb!]
\centering
    \includegraphics[width=\columnwidth]{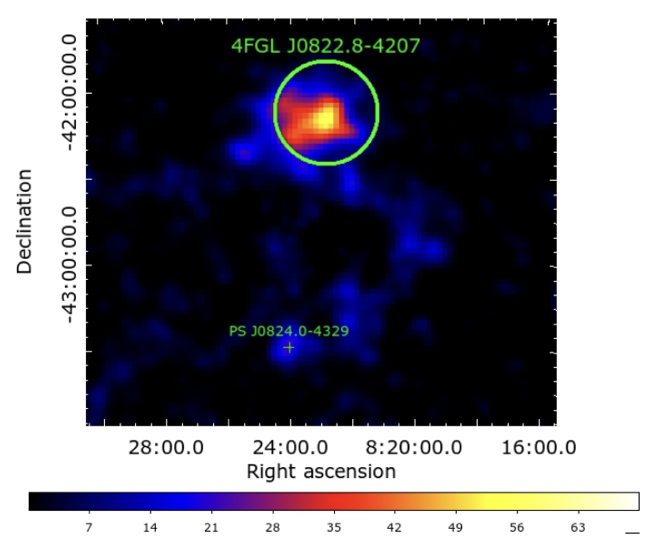}
    \caption{residual TS map showing the two $\gamma$-ray sources outside from Puppis A (4FGL J0822.8-4207 in the north a and PS J0824.0-4329 in the south.}
    \label{fig:sources}
\end{figure}

In order to find the best spatial model reproducing the morphology of Puppis A, we compared several templates (listed in Table \ref{tab:spatial_model}), following the method in \cite{hgl2012}\footnote{We notice that for the morphological analysis we investigate different spatial templates for Puppis A assuming its spectral shape a \texttt{LogParabola} (see Sect. \ref{sec:spec_puppisA})}.
The templates are compared by using the Akaike Information Criterion (AIC, \citealt{Akaike1998}) enabling the comparison of non-nested templates based on their AIC value, $AIC = -2\log(L) + 2k$, where $k$ is the number of degree of freedom\footnote{{For every source we count the spatial and spectral parameters. For the spatial coordinates the disk includes center and radius, the point like sources include their center. The spectral shape of Puppis A (Disk or image) includes always 3 parameters ($N_0, \alpha,\beta$, see Eq. \ref{eq:logp}), while pointlike sources have only 2 parameters ($N_0, \gamma$, see Eq. \ref{eq:powfermi})}} and L the likelihood. 
We consider a better fit when the $\Delta$AIC is higher, where $\Delta$AIC= AIC$_0$ - AIC$_{i}$ (AIC$_0$ is the AIC obtained for the disk plus two pointlike sources, taken as a reference, and AIC$_{i}$ is the AIC of the tested template.
The asymmetric $\gamma-$ray emission of Puppis A is confirmed by comparing the fit obtained with the uniform disk provided in the catalog 4FGL-DR3 with and without the two additional pointlike sources within the remnant (namely, PS J0823.2-4244 and PS J0823.8-4255). The likelihood way significantly improved by including the additional sources, further confirming that the uniform disk cannot describe the morphology of the remnant. 

Similar results are obtained by using radio maps of Puppis A as templates for the $\gamma-$ray morphology. In particular, we considered the \textit{Plank} maps at 33 GHz and 44 GHz, the \textit{SUMSS} map at 843 MHz and the \textit{VLA} map at 1.4 GHz. We show as example VLA contours in the middle panel of Fig. \ref{fig:Resid_disk}. We clearly find that none of these radio templates can properly describe the morphology of the observed $\gamma-$ray emission of Puppis A (specially in its eastern side), as shown by the likelihood values reported in Table \ref{tab:spatial_model}.

\begin{table*}[htb]
	\caption{Comparison between different spatial models used to fit the morphology of Puppis A between 1 GeV and 1 TeV.}
	\begin{center}
		\footnotesize
		\begin{tabular}{c c c c}
		    \hline\hline
			Spatial Model & Log Likelihood$^{a}$ & N d.o.f. & $\Delta$AIC\\
			\hline
            Disk + 2 pt & 2609.1 & 14 & 0\\
            Disk & 2471.5 & 6 & -259.1\\
            \textit{Planck} ($\nu = 33 GHz$) & 1838.1 & 3 &  -1519.9\\
            \textit{Planck} ($\nu = 44 GHz$) & 1815.6 & 3 & -1564.8\\
            \textit{SUMSS} ($\nu = 843 MHz$) & 2137.8 & 3 & -920.5\\
            \textit{VLA} ($\nu = 1.4 GHz$) & 2430.4 & 3 & -335.4\\
            %\textit{ROSAT} (0.1 - 2.0 keV) & 2624.0 & 3 & 51.9\\
            \textit{eROSITA} (0.2 - 0.7 keV) & 2608.7 & 3 & 21.4\\
            \textit{eROSITA} (0.7 - 1.0 keV) & 2613.6 & 3 & 31.0\\
            \textit{eROSITA} (1.0 - 8.0 keV) $^{b}$ & 2587.1 & 3 & 22.0\\
            \textit{XMM-Newton} (0.3 - 0.7 keV) & 2612.8 & 3 & 29.5 \\
            \textit{XMM-Newton} (0.7 - 1.0 keV) & 2618.9 & 3 & 41.7\\
            \textit{XMM-Newton} (1.0 - 8.0 keV) & 2600.1 & 3 & 4.1\\
            %\textit{XMM} (0.2 - 8.0 keV) & 2624.593 & 3 \\
            %N.A.$^a$ \textit{XMM} (0.2 - 0.7 keV)  & 2578.985 & 3 \\
            Un-absorbed \textit{XMM} (0.7 - 1.0 keV) & 2624.0 & 3 & 51.8\\
            %N.A.$^a$ \textit{eROSITA} (0.2 - 0.7 keV)  &  & 3 \\
            Un-absorbed \textit{eROSITA} (0.7 - 1.0 keV) & 2629.4 & 3 & 62.8\\
            \hline
            $^{a}${Difference between Log Likelihood with and without the source}\\
            $^{b}${Same spatial template as in the 4FGL-DR4 catalog.}
		\end{tabular}
        \label{tab:spatial_model}
	\end{center}
\end{table*}

On the other hand, the residual TS maps obtained with X-ray templates, i.e. the Puppis A maps produced from \textit{XMM-Newton} \citep{lsd16} and \textit{eROSITA} \citep{mbp22} observations\footnote{We do not consider \textit{ROSAT} observation as in \cite{hgl2012} because almost 20\% of the emission in the southern part of the remnant has not been detected, as shown in \cite{dlr13}.}, do not provide excess in the eastern part of the remnant and result in a better AIC (Table \ref{tab:spatial_model}). 
By comparing the likelihoods obtained with the different X-ray templates (see Table \ref{tab:spatial_model}), it is clear that the soft X-ray templates in the $0.7-1$ keV band (obtained both with \emph{XMM-Newton} and \emph{eROSITA}) provide a significant improvement with respect to those in the hard band ($1-8$ keV). 
Moreover, the soft X-ray band is a better diagnostic of the total density of the X-ray emitting plasma, since the hard band could miss the densest and coolest parts. The $\gamma$-ray emission depends on the total density, making the improvement of the fit remarkably coherent.
We warn the reader that the values for the very soft X-ray band (0.2-0.7 keV) should be taken with some caution, because of the higher absorption effects in that band. 

It must be also noted that the X-ray maps in the $0.7-1$ keV band can be affected by inhomogeneities in the interstellar absorption across the remnant\footnote{The effects of absorption are expected to be smaller in the hard X-rays.}. Significant variations in the interstellar column density, $N_H$, have been measured through spatially resolved X-ray spectral analysis \citep{dlr13,mbp22}, showing the highest values in the southern part of the shell. %, i. e., where we observe an excess in the $TS$ map (lower panel of Fig. \ref{fig:Resid}). 
To correct for this bias, we adopt as a template the un-absorbed X-ray map of Puppis A, which is obtained by following the approach described in \cite{mbp22}. The un-absorbed map of Puppis A in the $0.7-1$ keV band is shown in the left panel of Fig. \ref{fig:template_correct}. The right panel of Fig. \ref{fig:template_correct} shows the corresponding $TS$ map, which confirms that this template provides the best description of the $\gamma-$ray emission (see also Table \ref{tab:spatial_model})
%This effect can be easily corrected using the un-absorbed flux (Fig. \ref{fig:template_correct}). 

\begin{figure}[ht!]
    \centering
    \includegraphics[width=\columnwidth]{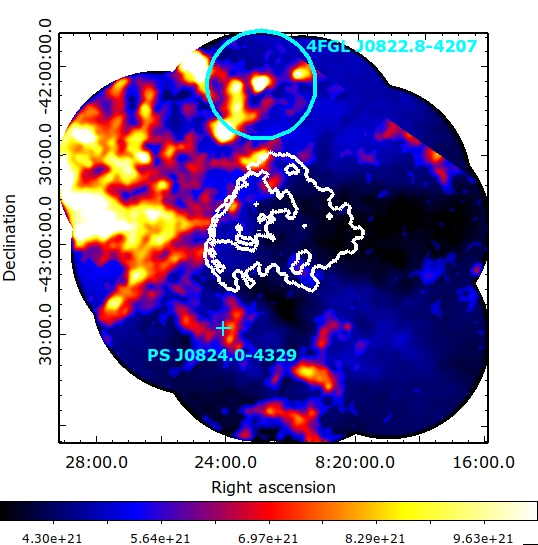}
    \caption{Map of the distributions of the ISM proton column densities, $N_p (H_2 + H_I)$, in the velocity range of 8-20 km s$^{-1}$ (see Fig. 8 in \citealt{asf22}). The white contours map the soft X-ray emission of Puppis A and the two cyan regions mark the two $\gamma$-ray excesses found outside from the remnant.}
    \label{fig:ism}
\end{figure}

The un-absorbed map in the $0.7-1$ keV band can be considered as a good proxy for the plasma emission measure, indicating denser material in the northeastern part of Puppis A and a more tenuous environment in the south-west. This is coherent with the distribution of the ambient density: dense molecular clouds in the north-east and more tenuous atomic clouds in the south-west \citep{asf22}. Similar results can be obtained by looking at the distribution of H in the region of Puppis A. Figure \ref{fig:ism} shows the distributions of the ISM proton column density, $N_p (H_2 + H_I)$, in the velocity range 8-20 km s$^{-1}$ (see Fig. 8 in \citealt{asf22}).

We then divided the unabsorbed X-ray template in two parts in order to analyze the two sides separately. In particular, we adopted a different strategy with respect to \cite{hgl2012}, dividing the template using a diagonal line, as shown in the left panel of Fig. \ref{fig:template_correct}, taking care of excluding all bright areas from the faint West part. This method provides a more precise division between the very bright and the dim X-ray emission (reflecting the regions with high$/$low ambient density, respectively), which is not taken into account in the methods in \cite{hgl2012}. 
This division does not improve significantly the fit (as in \citealt{hgl2012}), but it will provide important information for the spectral analysis (see Section \ref{sec:spec_puppisA}). 

In conclusion, by comparing all the templates with the AIC, the best model which reproduces the morphology of Puppis A is the X-ray template with the un-absorbed \textit{eROSITA} emission in the 0.7-1.0 keV energy (Fig. \ref{fig:template_correct}).

\subsection{Detection of nearby sources} 
\label{sect:sources}
Using the \textit{eROSITA} un-absorbed map in the 0.7 - 1 keV band as a template for the morphology of Puppis A, we then analyzed the source in the north of the remnant (4FGL J0822.8-4207, \citealt{agk2022}) and that in the south (namely, PS J0824.0-4329). %OK

We confirm the detection of 4FGL J0822.8-4207, which was first detected in \cite{bbd20} and deeply studied in \cite{agk2022} and classified as a point-like source. To describe the morphology of the source and assess its angular extension (if any), we adopted a point-like template, a radial Gaussian template and a uniform disk template. For this analysis we let free to vary also the center of the source. we found $TS_{ext}$= 22.0, where TS$_{ext}$ = 2 $\ln L_{ext}/L_{ps}$ and $L_{ext}$ and $L_{ps}$ are the likelihoods for the fit with the extended (Gaussian or uniform disk) and the point-like source, respectively. Values of $TS_{ext}>16$ provide a statistically significant detection of an extended source \citep{laa2012}, so we conclude that 4FGL J0822.8-4207 cannot be considered as a point-like source. 
Very similar results are obtained with the uniform disk template ($TS_{ext} = 18$).

In conclusion, the source 4FGL J0822.8-4207 is significantly extended and its morphology can be modelled as a radial Gaussian with sigma of $0.15 \pm 0.03^\circ$ or, equivalently, as a disk with radius of $0.30 \pm 0.05^\circ$.
Figure \ref{fig:sources} and \ref{fig:ism} show the position of 4FGL J0822.8-4207 and its angular extension in the uniform disk scenario.

The same analysis has been performed for the new putative source we detected with our analysis, namely PS J0824.0-4329 (see Sect \ref{sec:morph} and Table \ref{tab:PS}). This source is located beyond the southern border of Puppis A (right panel of Fig. \ref{fig:Resid_disk} and Fig. \ref{fig:ism}). In this case, the TS$_{ext}$ = 7.6 does not reach the minimum requested value of 16 for a source to be considered as extended, so PS J0824.0-4329 is consistent with being a point-like source. 
Figure \ref{fig:sources} shows the two $\gamma$-ray sources superimposed with their spatial model. For visualization purposes we used the uniform disk for 4FGL J0822.8-4207.

\subsection{Spectral analysis}
\label{sec:spec_puppisA}
We here show the spectral analysis for all the sources studied in this paper. Since the aim of the first part of the paper is to emphasize differences between the two sides of the remnant, if any, we conducted the spectral analysis in the 300 MeV - 1 TeV energy band.
As discussed in Sect. \ref{sec:morph} we used the un-absorbed \textit{eROSITA} image in the $0.7-1$ keV  band as the spatial template for the morphology of Puppis A (left panel of Fig. \ref{fig:template_correct}) and the radial Gaussian with sigma of $0.15 \pm 0.03^\circ$ for 4FGL J0822.8-4207. All the additional point-like sources out of the remnant found in Section \ref{sec:morph} are incorporated in the template, including PS J0824.0-4329.

The spectrum of Puppis A has been fitted by adopting the spectral model \texttt{Log Parabola} available in \texttt{gtlike}:

\begin{equation}
\label{eq:logp}
    \frac{dN}{dE} = N_0 \left(\frac{E}{E_b}\right)^{-(\alpha + \beta \log(E/E_b))}.
\end{equation}
where $N_0$ is the differential photon flux at $E_b$, $\alpha$ and $\beta$ are the photon index at $E_b$ and the half-curvature of the LogParabola, respectively, and $E_b$ is a scale parameter, fixed at 1 GeV. 

\begin{figure}
\centering
    \includegraphics[width=\columnwidth]{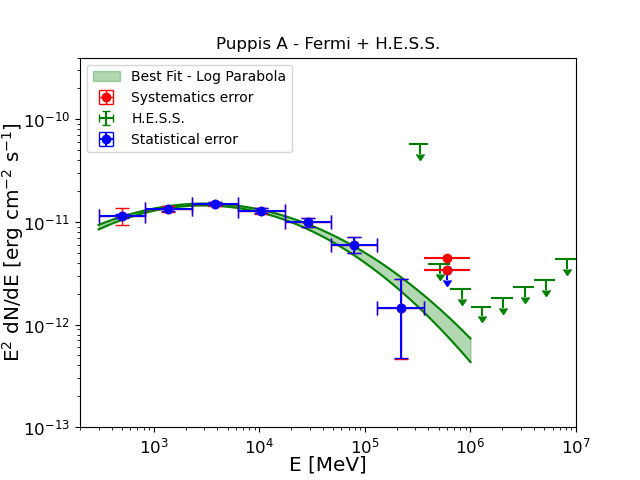}
    \caption{SED of Puppis A with the corresponding best fit model and its uncertainties at 68\% confidence level (green curves and shaded region). Fermi upper limits are at 95\% confidence level. Statistical and systematic errors are indicated in blue and red, respectively. Systematics errors for the Fermi upper limits present the extrema. The green arrows show the H.E.S.S. upper limits \citep{HessColl15}.}
    \label{fig:fermi+hess}
\end{figure}

Thanks to the improved statistics in the data analyzed here, we find that this model provides a significantly better description ($\Delta(TS)=22$) of the \emph{Fermi}-LAT data points than a simple power law, which was adopted by \citet{hgl2012}. The SED of Puppis A is shown in Fig. \ref{fig:fermi+hess}, where the blue crosses mark the SED we produced with the \textit{Fermi}-LAT data (with the statistical errors), while the green arrows mark the upper limits obtained with H.E.S.S. \citep{HessColl15}. 

\begin{figure}[h!]
\centering
    \includegraphics[width=\columnwidth]{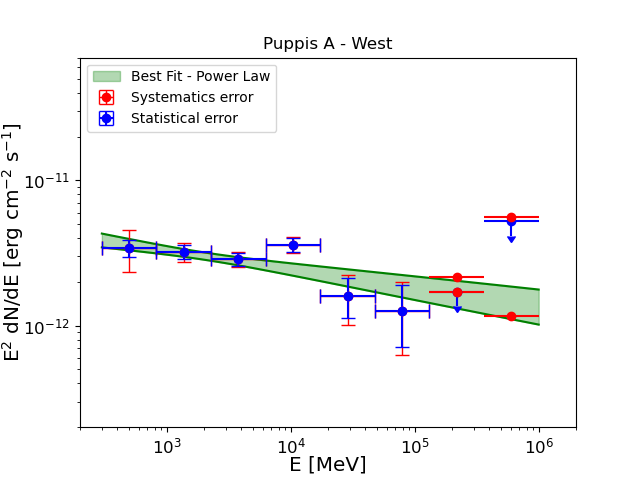}
    \includegraphics[width=\columnwidth]{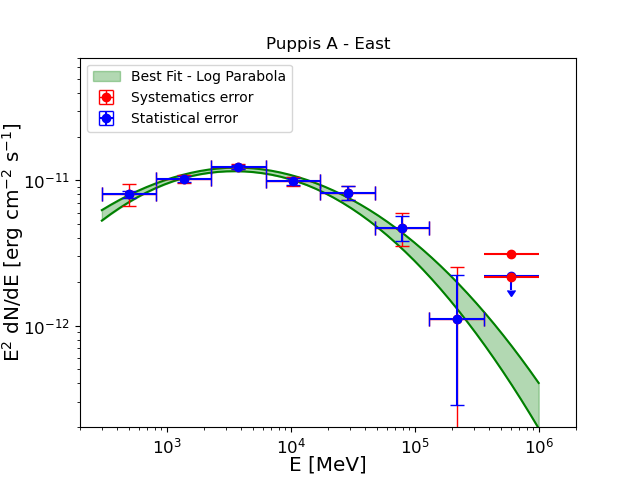}
    \caption{\emph{Upper panel:} SED of the southwestern side of Puppis A, with the corresponding best fit model and its uncertainties at 68\% confidence level (green curves and shaded region). Fermi upper limits are at 95\% confidence level. Statistical and systematic errors are indicated in blue and red, respectively.  Systematics error for the Fermi upper limits present the extrema. \emph{Lower panel:} same as upper panel for the northeastern side.}
    \label{fig:East_West}
\end{figure}

In order to analyze the spectra of the northeastern and southwestern sides separately, we divided the template in two parts, as discussed in Section \ref{sec:morph} and shown in Fig. \ref{fig:template_correct}. 
As expected, the whole Puppis A, its northeastern part and its southwestern part are all significantly above the background, with a $TS$ value of 5429 for the whole Puppis A, 3555 for its northeastern side, and 390 for its southwestern side. The upper and lower panels of Fig. \ref{fig:East_West} show the SED that we obtained for  the southwestern and northeastern part of Puppis A, respectively. While the smaller statistics of the data analyzed by \cite{hgl2012,xgl17} hampered the possibility of finding different spectral shapes between the two halves of the remnant, we find that the two spectra are different. In particular, the high statistics available allows us to find that the \texttt{Log Parabola} model fits the substantial curvature observed in the northeastern SED significantly better than a simple power law ($\Delta(TS) = 88$). On the other hand, we find that the spectral curvature is not significant up to the $3 \sigma$ level for the southwestern side, so we fitted the spectrum with a power law:

\begin{equation}
\label{eq:powfermi}
    \frac{dN}{dE} = N_0 \left(\frac{E}{E_0}\right)^{-\gamma},
\end{equation}
where $N_0$ is the differential photon flux at $E_0$ (see Table \ref{tab:fit_pupa}), $\gamma$ is the photon index and $E_0$ is the energy scale, set to 1 GeV.
Our results are then consistent with the model found in \cite{hgl2012,xgl17} for the southwestern side.

The best fit parameters  with the associated errors  are shown in Table \ref{tab:fit_pupa}. For each parameter in the Table, the first error corresponds to the statistical error, while the second one indicates the systematic error, obtained by following the prescription
developed in \citet{pbj13, aaa16}. In the  systematic errors we also included the uncertainties associated with the effective area\footnote{\url{https://fermi.gsfc.nasa.gov/ssc/data/analysis/LAT_caveats.html}}. 
%Our analysis significantly improves the results in \cite{hgl2012}, showing that a \texttt{Log Parabola} model describes the data points of the northeastern side significantly better than a power law, thus being able to reveal the curvature of the spectrum. 
%On the other hand, our results are consistent with the model found in \cite{hgl2012,xgl17} for the southwestern side.
The fluxes in the $300$ MeV $-$ 1 TeV energy band are $F_{NE}=(5.13 \pm 0.08) \times 10^{-11}$ erg cm$^{-2}$ s$^{-1}$ and $F_{SW}=(1.87 \pm 0.035) \times 10^{-11}$ erg cm$^{-2}$ s$^{-1}$ for the northeastern and southwestern side, respectively.

\begin{table*}[h!]
	\caption{Best fit values for the spectral analysis in the 300 MeV - 1 TeV energy band. For each parameter, the first and second error correspond to the statistical and systematic errors, respectively.}
	\begin{center}
		\footnotesize
		\begin{tabular}{c c c c c}
		    \hline\hline
			Source & Differential photon flux at $E_0^{(a)}$ (MeV cm$^{-2}$ s$^{-1}$) & Photon Index & alpha & beta \\
			\hline
            4FGL J0822.8-4207 & $(2.6 \pm 0.3 \pm 0.5) \times 10^{-8}$& 2.05 $\pm$ 0.08 $\pm$ 0.09 & - & - \\
            PS J0824.0-4329 & $(7 \pm 6 \pm 3) \times 10^{-8}$& 1.8 $\pm$ 0.2 $\pm$ 0.1& - & - \\
            Puppis A & $(8.2 \pm 0.2 \pm 0.9) \times 10^{-6}$ & - & 1.80 $\pm$ 0.03 $\pm$ 0.08 & 0.10 $\pm$ 0.01 $\pm$ 0.02 \\
            Puppis A north-east & $(6.1 \pm 0.2 \pm 0.7)\times 10^{-6}$ & - & 1.72 $\pm$ 0.03 $\pm$ 0.07 & 0.12 $\pm$ 0.01 $\pm$ 0.02\\
            Puppis A south-west & $(1.6 \pm 0.2 \pm 0.1)\times 10^{-6}$ & 2.14 $\pm$ 0.05 $\pm$ 0.05 & - & - \\
            \hline
            $^{a}${Energy scale set to 1 GeV}
		\end{tabular}
		\label{tab:fit_pupa}
	\end{center}
\end{table*}

\begin{figure}
\centering
    \includegraphics[width=.49\textwidth]{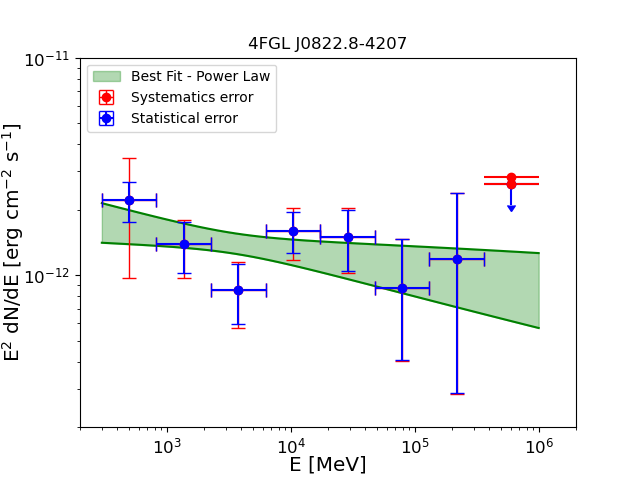}
    \includegraphics[width=.49\textwidth]{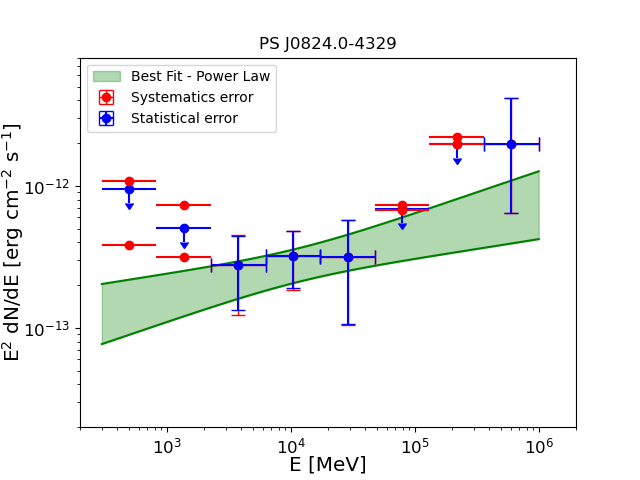}
    \caption{\emph{Upper panel:} SED of the source 4FGL J0822.8-4207 with the corresponding best fit model and its uncertainties at 68\% confidence level (green curves and shaded region). Fermi upper limits are at 95\% confidence level. Statistical and systematic errors are indicated in blue and red, respectively.  Systematics error for the Fermi upper limits present the extrema. \emph{Lower panel:} same as upper panel for the source PS J0824.0-4329.}
    \label{fig:sed_j0822.8}
\end{figure}

We also analyzed the spectral shapes of the sources 4FGL J0822.8-4207 and PS J0824.0-4329. The $\gamma-$ray emission of both sources is significantly above the background, with $TS=99$ for 4FGL J0822.8-4207 and $TS=23$ for PS J0824.0-4329.
In both cases, the SED can be fitted by a power law (Fig. \ref{fig:sed_j0822.8}). All the best fit parameters, together with the flux in the 300 MeV - 1 TeV energy range, are summarized in Tab. \ref{tab:fit_pupa}. Results for the source 4FGL J0822.8-4207 are consistent with those obtained in \citealt{agk2022}.

%In both cases, the SED can be fitted by a power law, with photon index $\Gamma = 2.0 \pm 0.08 \pm 0.09$\footnote{First and second error correspond to the statistical and systematic errors, respectively} for 4FGL J0822.8-4207 (consistent with that obtained by \citealt{agk2022}) and $\Gamma = 1.8 \pm 0.2 \pm 0.1$ for PS J0824.0-4329. Figure \ref{fig:sed_j0822.8} shows the SED of 4FGL J0822.8-4207 (upper panel) and of PS J0824.0-4329 (lower panel). Best fit parameters are reported in Table \ref{tab:fit_pupa}. 
%The fluxes in the $300$ MeV $-$ 1 TeV energy band are $(9.4 \pm 1.4) \times 10^{-12}$ erg cm$^{-2}$ s$^{-1}$ and $(2.7 \pm 1.3) \times 10^{-12}$ erg cm$^{-2}$ s$^{-1}$ for 4FGL J0822.8-4207 and PS J0824.0-4329, respectively.

\section{Discussion}
\label{sec:discussion}

The morphological analysis described in Section \ref{sec:morph} has revealed a clear asymmetry in the $\gamma$-ray emission of Puppis A: the northeastern part of the remnant, characterized by the brightest X-ray emission and by an enhanced ambient density, also shows the highest $\gamma$-ray emission. 
The asymmetry in the morphology of the source seems to reflect also a difference in the spectral shape of the two sides of the remnant, with the northeastern part showing a curved spectrum, which can be fitted by the \texttt{Log Parabola} model, and the southwestern part showing a SED which can be nicely fitted by a power law (see Table \ref{tab:fit_pupa}). This is suggestive of a possible difference in the physical origin of the $\gamma-$ray emission.

Moreover, we studied two sources located outside from the remnant: the northern source 4FGL J0822.8-4207 (already detected by \citealt{bbd20,agk2022}) and the southern source PS J0824.0-4329. Remarkably, the position of each of these two sources corresponds to an enhancement in the ambient density, as shown in Fig. \ref{fig:ism}.
The two sources are characterized by a hard spectrum, which can be fitted by a power law (see Table \ref{tab:fit_pupa}).  

We can divide the results of this project in two main categories: the physical origin of the $\gamma$-ray emission from Puppis A and the physical origin of the two sources outside from the remnant. We will discuss them separately.

\subsection{Physical origin of the $\gamma$-ray emission of Puppis A}
\label{sec:discpuppis}

As explained above, Puppis A evolves in a very inhomogeneous environment: while the ambient density is relatively low in the southwestern part of the shell, denser material and molecular clouds are observed at the northeastern edge of the remnant (see \citealt{asf22} and Fig. \ref{fig:ism}). It is then natural to speculate that the different properties of the $\gamma-$ray emission in these two parts of the remnant can be the result of the propagation of the shock in different environments. 

As previously suggested in \citealt{hgl2012,xgl17}, the leptonic scenario as origin of the $\gamma$-ray emission in Puppis A is strongly disfavored. Indeed, leptonic cases require an electron-to-proton ratio at least ten times higher than the standard CR abundance. The IC-dominated model necessitates an extremely low density along with a relatively weak magnetic field, which is definitely not the case for Puppis A. Additionally, the two spectral shapes (Fig. \ref{fig:East_West}) suggest an hadronic scenario on both sides of the remnant.

In this framework, one can think of two different populations of CRs (one at the southwest and one at the northeast of the remnant) associated with different acceleration mechanisms:
\begin{itemize}
    \item \textit{Southwestern part:} The southwestern side of Puppis A has been found to be interacting with a relatively tenuous atomic cloud \citep{asf22}, which is in agreement with the faint X-ray and $\gamma$-ray emission observed therein. Moreover, from the analysis of \emph{eROSITA} data, it has been found that the particle density of the X-ray emitting plasma in this region is the lowest in the remnant, being of the order of $\sim 1$ cm$^{-3}$ \citep{mbp22}. The propagation of the southwestern shock in such a low density environment makes the remnant dynamically young, allowing the shock to accelerate particles via DSA.
    In this framework, one can assume a power law spectrum for the energy of the accelerated protons ($E^{-2}$). We then synthesized the  hadronic $\gamma$-ray emission \citep{kat14} from this distribution of protons by also including an exponential cutoff at 1 TeV in the hadronic spectrum (see upper panel of Fig. \ref{fig:sed_east_west_reacc}). Results do not change significantly provided that the cutoff is  $<5$ TeV. Such a low cut-off energy for the particle spectrum is in agreement with what was previously found in \cite{HessColl15}.

    The resulting $\gamma$-ray flux can be written as \citep{gac09}:
\begin{equation}
\label{eq:Fgamma}
    F_\gamma = 2 \times 10^{-4} \left (\frac{E_{CR}}{10^{50}~\rm{erg}} \right) \left(\frac{n}{\rm{cm}^{-3}} \right) \left(\frac{D}{\rm{kpc}} \right)^{-2} \; \rm{MeV} \; \rm{cm}^{-2} \; \rm{s}^{-1}
\end{equation}
Considering that the $\gamma-$ray flux we measured in the southwestern side is $F_{SW}=(1.87 \pm 0.035) \times 10^{-11}$ erg cm$^{-2}$ s$^{-1}$, assuming a distance $D=1.3$ kpc \citep{rcw17} and an ambient density  $n=1$ cm$^{-3}$ \citep{mbp22}, the CR energy is $E_{CR}\sim10^{49}$ erg.
    The comparison between the data and the model is shown in the upper panel of Fig. \ref{fig:sed_east_west_reacc}. This scenario (hereafter ``DSA model") provides a very good description of the observed SED and a very reasonable value for the energy in the local cosmic rays.
    \\
    \item \textit{Northeastern part}. \cite{asf22} found that the northeastern part of Puppis A is interacting with a dense and massive molecular cloud. Because of the proportionality of the shock velocity with the inverse of the square root of the ambient density, in this region the shock is expected to be significantly slower than in the southwest. The low velocity of the shock propagating in a dense cloud ($V_s < 200$ km s$^{-1}$) can make it radiative, thus inducing a rapid compression of the shocked cloud material. \cite{bc82} demonstrated that the re-acceleration of pre-existing CR electrons at a cloud shock and subsequent adiabatic compression results in enhanced synchrotron radiation, capable of explaining the radio intensity from evolved SNRs. Re-acceleration of pre-existing CRs, and the subsequent $\gamma-$ray emission via proton-proton collision with the compressed post-shock material, was then adapted by \cite{ubf10} to reproduce the $\gamma-$ray spectral properties of interacting remnants. This first scenario can explain the gamma-ray spectrum derived in our analysis in the northeastern side of Puppis A: the $\gamma$-ray emission, via proton-proton collision with the compressed post-shock material, is then enhanced by the adiabatic compression, to very high densities, of the plasma behind the shock. Moreover, the compression of pre-existing CRs in the radiative shell enhances the CR spectrum both energizing particles and increasing the normalization of the spectrum. The resulting spectrum of compressed CRs is shown below \citep{ubf10}.
    \begin{equation}
        n_{comp}(p) = \xi^{2/3} n_{GCR} (\xi^{1/3} p)
    \end{equation}
    where $n_{GCR}(p)$ is the density of Galactic cosmic rays, $p$ is the momentum, $\xi$ is the adiabatic compression ratio ($n_{shell}/n_0r$, where r is the shock compression ratio and $n_{shell}$ and $n_0$ are the gas density in the shell and in the ambient medium, respectively).
    The pre-existing CRs can also be reaccelerated at the shock front. Then, in shock-cloud interaction this mechanism can enhance the $\gamma$-ray emission. Finally, only pre-existing CRs can be responsible of $\gamma$-ray emission in SNRs interacting with MCs without invoking particle acceleration at the shock.

    Several parameters can affect the resulting $\gamma-$ray emission, we here adopt those listed in Table \ref{tab:reacc} to synthesize the expected SED in this scenario (hereafter reacceleration model). A crucial parameter is the upstream cloud density, which we set  to $250$ cm$^{-3}$ on the basis of the analysis by \cite{asf22}. The value of the cloud filling factor reported in Table \ref{tab:reacc} refers to the shocked cloud only (this explains its low value). While distance, age and radius were set according to what was reported in the literature (see Section \ref{sec:intro}), the other (still unknown) parameters were set to ad-hoc, reasonable, values. The lower panel of Fig. \ref{fig:sed_east_west_reacc} shows the comparison between the observed SED and the reacceleration model. The model nicely fits the data points, very well reproducing the observed curvature of the SED. On the other hand, one can note that the proton cutoff energy needed to reproduce the northeastern data (600 GeV, see Table \ref{tab:reacc}) is only a factor of a few smaller than the cutoff energy of the protons in the DSA model adopted at southwest ($1-5$ TeV).
    \\
    However, we caution the reader that the model in \cite{ubf10} cannot be applied for too low shock velocities. Through CO and HI measurements, \cite{asf22} found shock velocity of around 10 km s$^{-1}$ which would not suit with the re-acceleration model described above. In this case the most suitable approach involves particle acceleration driven by the reflected shock, as in \cite{iyi12} for the supernova remnants RX J1713.7-3946. In this scenario, considering a shock velocity of around 10 km s$^{-1}$ and a cloud average density of 250 cm$^{-3}$, the density of the diffuse gas required to drive this shock into the cloud (for shock velocities around 1000 km s$^{-1}$) is around 0.02 cm$^{-3}$, much lower than what was estimated through the X-ray emission (n $\sim$ 1 cm$^{-3}$ in \citealt{mbp22}).

    \begin{figure}
    \centering
    \includegraphics[width=.49\textwidth]{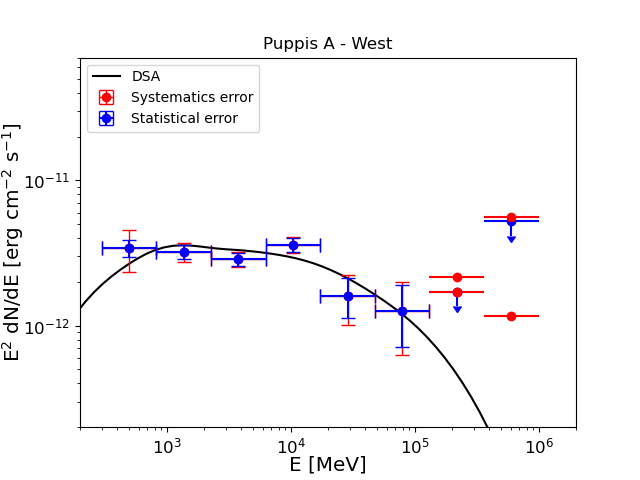}
    \includegraphics[width=.49\textwidth]{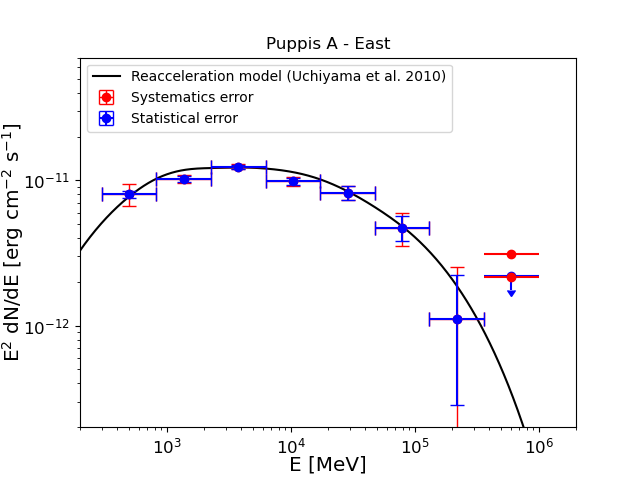}
    \caption{\emph{Upper panel}: SED of the southwestern side of Puppis A (same as the upper panel in Fig. \ref{fig:East_West}). The black curve shows the hadronic emission expected from a power law distribution of proton energies ($E^{-2}$) with a cutoff at 1 TeV. \emph{Lower panel}: SED of the northeastern side of Puppis A (same as the lower panel in Fig. \ref{fig:East_West}). The black curve shows the hadronic emission expected from the reacceleration model (see Table \ref{tab:reacc} for the model parameters).}
    \label{fig:sed_east_west_reacc}
    \end{figure}
\end{itemize}

In order to have a confirmation of the fitting models, realistic models that also include the contribution, e.g., from primary and secondary electrons radiation, with a comparison with the radio data, needs to be analysed for both acceleration and re-acceleration model.

An alternative scenario would consist in invoking the DSA model for both sides of the remnant. 
The shapes of the two SEDs shown in Fig. \ref{fig:sed_east_west_reacc} look similar, so one may be tempted to fit also the northeastern SED with a DSA model. 
By assuming the same values of $E_{CR}$ for the two sides, the northeast to southwest flux ratio ($F_{NE}/F_{SW}=f\approx2.8$) might be explained by a similar ratio $f$ in the ambient densities, with the density at northeast being only a factor $<3$ larger than that at southwest. Such a difference in the ambient density can be related to a similar ratio of absorption on the line of sight, in agreement with \cite{asf22}. We note that the density contrast of the X-ray emitting plasma between the northeastern and the southwestern side of Puppis A is of the order of $n_{NE}/n_{SW}\approx 4-5$ \citep{mbp22}, which is not that far from $f$.
According to Eq. \ref{eq:Fgamma}, with a much higher particle density, an unrealistically low value for the cosmic ray energy at northeast, $E_{CR}\sim10^{47}$ erg, would be necessary to explain the observed flux $F_{NE}$ with the DSA model. This CR energy looks indeed too low for particles accelerated via DSA, and is also smaller than that obtained in the southwestern side. 
However, cosmic-ray energies of around $10^{47}$ erg in SNRs has been previously found in \cite{asf22,fsy21}, explaining that both the filling factor and the CR escape outside of the shell can make the cosmic-ray energy much smaller than $10^{49}$.
Words of caution are needed when assuming that the entire northeastern shock front is interacting with the cloud. Indeed, the fraction of the shock interacting with the cloud is still uncertain.
Protons in the HI medium and protons in the low density ambient medium can both act as targets in the proton-proton reaction. If we assume that both are located in a similar position within the SNR, the high density will enhance the gamma-ray emission: this situation can occur if the two gases with different densities have complementary distribution. However, the presence of HI clumps embedded in tenuous hot gas cannot be resolved at the current HI resolution.
In conclusion, the scenario which considers canonical DSA in both sides of the remnant presents some issues, namely: i) it does not provide realistic values of the CR energy, unless ignoring the shock-cloud interaction observed by \citet{asf22}; ii) even excluding shock-cloud interaction, it does not perfectly explain the differences in $\gamma-$ray fluxes between the northeastern and southwestern side in terms of the density contrast observed in X-rays. However, this latter scenario, though less likely, cannot be definitely ruled out at the moment.

\begin{table}[h!]
	\caption{Parameters used to set the crushed cloud reacceleration model by \cite{ubf10} for Puppis A.}
	\begin{center}
		\footnotesize
		\begin{tabular}{c c}
		    \hline\hline
			Input Parameter & value  \\
			\hline
		    Distance & 1.3 kpc\\
            Age & 4 kyr\\
            Radius & 16 pc\\
            Filling Factor & 0.01\\
            Upstream magnetic field & 30 $\mu$G \\
            Upstream plasma density & 250 cm$^{-3}$ \\
            Shock velocity & 50 km s$^{-1}$\\
            p$_{max}$ & 600 GeV c$^{-1}$\\
            \hline
            Output Parameter & value \\
            \hline
            Radiative compression parameter & 5 \\
            Downstream magnetic field & 500 $\mu$G \\
            Downstream plasma density & $5 \times 10^{3}$ cm$^{-3}$ \\
            \hline
		\end{tabular}
		\label{tab:reacc}
	\end{center}
\end{table}

\subsection{Physical origin of the $\gamma$-ray emission from 4FGL J0822.8-4207 and PS J0824.0-4329}

\begin{figure}
\centering
\includegraphics[width=\columnwidth]{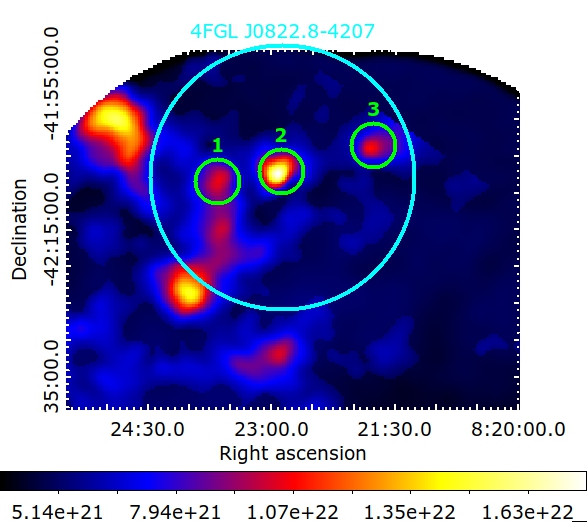}
\includegraphics[width=\columnwidth]{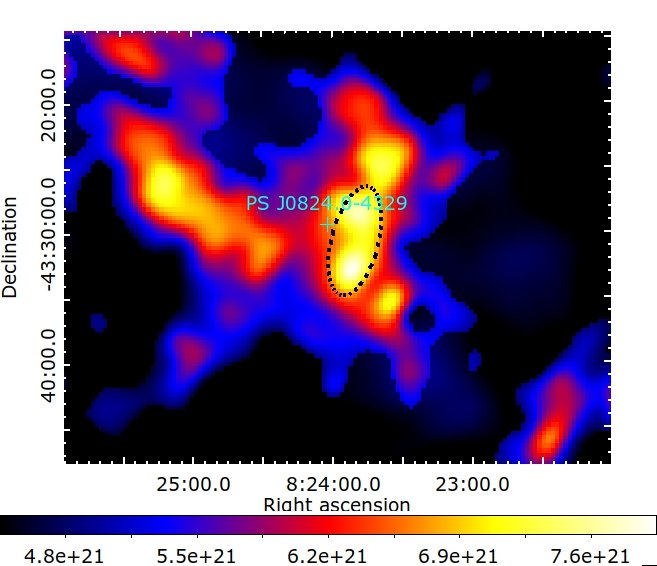}
\caption{Close-up view of the map of the distributions of the ISM proton column densities, $N_p (H_2 + H_I)$, in Fig. \ref{fig:ism},centered at 4FGL J0822.8-4207 (\emph{upper panel}) and PS J0824.0-4329 (\emph{lower panel}). Green regions in the upper panel and the dashed black region in the lower panel mark the region used to measure the volume of the interacting cloud.}
\label{fig:zoom_ism}
\end{figure}

Figure \ref{fig:zoom_ism} shows a close-up view of the region of 4FGL J0822.8-4207 (upper panel) and PS J0824.0-4329 (lower panel) in the map of the ISM proton column density extracted in the velocity range 8-20 km s$^{-1}$, which corresponds to the distance of Puppis A (see also Fig. \ref{fig:ism} and Fig. 8 in \citealt{asf22}). The figure clearly reveals that both sources are associated with dense clumps of interstellar medium.
These two $\gamma$-ray excesses found outside of the remnant are then interesting sources, whose emission can be explained as the result of CRs accelerated and escaped from the remnant, now diffusing in nearby molecular clouds.
This scenario was invoked by \cite{agk2022} for 4FGL J0822.8-4207 (though an alternative explanation, considering the Herbig Haro object  HH219 as the origin of the GeV source was also proposed). We stress that, at odds with \cite{agk2022} who classified 4FGL J0822.8-4207 as a point-like source, we found that the source is extended (see Section \ref{sect:sources}). We here discuss whether the $\gamma-$ray emission of both 4FGL J0822.8-4207 and of the previously undetected putative source PS J0824.0-4329 can indeed be associated with CRs diffusing in nearby clouds.

We estimate the diffusion length of CRs with energy $E$ escaped from Puppis A and moving in the local ISM with magnetic field $B$ as $l_d=2\sqrt{D\tau_{SNR}}$, where the diffusion coefficient, $D$, can be calculated as shown in the following Equation \citep{gac09}:
\begin{equation}
\label{eq:diff}
D=3\times10^{27}(E/1~\rm{GeV})^{1/2}(B/3\rm{\mu G})^{-1/2}
\end{equation}
and $\tau_{SNR}=4$ kyr is the age of Puppis A. Assuming a typical value for the interstellar magnetic field, $B=3~\mu$G, and considering CRs with energy $E=10$ GeV, one obtains $l_d\sim 22$ pc. The projected distance between 4FGL J0822.8-4207 and the center of Puppis A is $\sim19$ pc, while that of PS J0824.0-4329 is $\sim16$ pc. Both $\gamma$-ray sources (and the dense clouds) are then located within $l_d$, assuming no offset along the line of sight, so CRs escaped from Puppis A can reach them. It is then possible that the observed $\gamma-$ray emission is associated with proton-proton collisions between the CRs escaped from the remnant and the dense cloud material.

In this scenario, one can estimate the energy density of the CRs inside the clouds from the $\gamma-$ray fluxes observed for the two sources and reported in Sect. \ref{sect:sources}. To this end, we here assume that the CRs escaped from Puppis A are uniformly distributed in a sphere with radius $l_d$, with an energy density $w_{CR}=W_{CR}/(\frac{4}{3}\pi l_d^3)$ (where $W_{CR}$ is the CR energy). 
For a cloud with volume $V_{cl}$ and density $n_{cl}$, the $\gamma-$ray luminosity can be estimated as \citep{gac09}:
\begin{equation}
L_\gamma=\frac{w_{CR}V_{cl}}{3\tau_{pp}}
\label{eq:lcloud}
\end{equation}  
where $\tau_{pp}$ is the mean time interval between proton-proton collisions and scales as $1/n_{cl}$ ($\tau_{pp} = 1/(n_{gas} \sigma_{pp} k c)$, where $\sigma_{pp} = 4 \times 10^{-26}$ cm$^2$, k=0.45 and $c$ is the speed of light.
To estimate the values of $V_{cl}$ and $n_{cl}$, we proceed as described below. We consider the cloud associated with 4FGL J0822.8-4207 as composed by three spherical clouds, whose projected size in the plane of the sky is indicated by the three circles ($3'$ in radius each) shown in the upper panel of Fig. \ref{fig:zoom_ism}. The radius of each circle corresponds to $1.1$ pc at 1.3 kpc. Within each circle, we then calculate the excess in the proton column density with respect to the ``background" value, which is $3\times10^{21}$ cm$^{-2}$, and derive the proton density accordingly. We find $n_{cl1}\sim900$ cm$^{-3}$, $n_{cl2}\sim2000$ cm$^{-3}$, and $n_{cl3}\sim1400$ cm$^{-3}$ for cloud 1, 2, 3 of Fig. \ref{fig:zoom_ism}, respectively. By including these values in Eq. \ref{eq:lcloud} and considering the $\gamma-$ray luminosity of 4FGL J0822.8-4207, we obtain an energy density $w_{CR}\sim 7$ eV cm$^{-3}$. We then adopt the same methodology for PS J0824.0-4329, by approximating the cloud with an ellipsoid, whose projected shape in the plane of the sky is shown in the lower panel of Fig. \ref{fig:zoom_ism} (with semi-axis $a=1.62$ pc, $b=c=0.68$ pc). We then obtain $n_{cl}\sim1900$ cm$^{-3}$ and  $w_{CR}\sim 7$ eV cm$^{-3}$. Remarkably, we find the same values of $w_{CR}$ in the two sources, which are located at opposite sides of Puppis A. In this scenario, one obtains $W_{CR}\sim1.5\times10^{49}$ erg, which is pretty similar to the value obtained in the analysis of the Puppis A SED (see Sect. \ref{sec:discussion}). We caution the reader that $E_{CR}$ and $W_{CR}$ are the CR energies at the source and at the clouds, respectively, and they do not necessarily need to have the same value.

Moreover, the value of the energy density of cosmic rays around Puppis A is well above the average value in the Milky-Way ($\overline{w_{CR}}$ = 1.8 eV/cm$^3$ \citealt{Webber1998}). This result strongly indicates that the $\gamma$-ray emission of 4FGL J0822.8-4207 and PS J0824.0-4329  originate from CRs escaped from the closely SNR Puppis A and interacting with dense clouds. 
If these sources are indeed linked to Puppis A, then their spatial distribution tells us that the escape of cosmic rays from the shell of Puppis A occurs anisotropically along the magnetic field lines. Indeed we would expect bright $\gamma$-ray sources in regions with high ambient density, toward the Northeastern part of Puppis A. The non-detection of $\gamma$-ray sources in that direction tends to indicate that the B field around Puppis A is preferentially oriented North-South.

\section{Conclusions}
\label{sec:conc}
This paper is focused on the study of the $\gamma$-ray emission stemming from the galactic SNR Puppis A and its surrounding area. We present the analysis of new \textit{Fermi}-LAT observations taking advantage of the high statistics available.
 
We confirm the asymmetric $\gamma$-ray emission of Puppis A, characterized by a bright northeastern side and a dim southwestern side, as previously found in \citealt{hgl2012,xgl17}. We find that the X-ray morphology, especially the \emph{eROSITA} map in the $0.7-1$ keV band, when corrected for the interstellar absorption, provides the best spatial template to reproduce the $\gamma$-ray map (see Table \ref{tab:spatial_model}).
Thanks to the high quality of the new data, we were able to perform a spatially resolved spectral analysis, which led to a hint of a difference in the spectral shape between the northeastern and southwestern parts of the remnant. In particular, the new data shows for the first time a significant curvature of the spectrum of the northeastern side of Puppis A.
Since this difference is indicative of two different population of CRs accelerated in the two regions, we investigate whether this can be the result of two different acceleration mechanisms being at work in different parts of Puppis A. 
Motivated by the strong indications of interaction between the northeastern part of the remnant and a massive molecular cloud recently found by \cite{asf22}, we explore the possibility of reacceleration of ambient cosmic rays in the radiative shock propagating into the cloud, by adopting the model proposed by \cite{ubf10}, and the scenario of reflected shocks in young SNRs, proposed by \cite{iyi12}. 
Despite the reacceleration model perfectly reproduces the observed SED of the northeastern part of Puppis A, it is in contrast with the low shock velocities measured in \cite{asf22} which would request a different approach, invoking particle acceleration caused by the reflected shock \citep{iyi12}. Nevertheless, this latter scenario would imply the presence of X-ray synchrotron emission, which has never been detected in Puppis A. Therefore, the question of the mechanism governing particle acceleration in the northeastern part of the remnant is still uncertain, leaving lots of open questions on supernova remnant interacting with molecular clouds. However, as shown in Sect. \ref{sec:discpuppis}, we suggest that the physical mechanism responsible for particle acceleration in the northeastern part of Puppis A needs to be different from the canonical DSA which perfectly reproduces the $\gamma$-ray emission from the southwestern side. Indeed, it evolves in a much more tenuous environment and its (dim) $\gamma-$ray emission can be naturally explained as hadronic emission from particles accelerated via DSA. 
This scenario considering two different CR populations between southwestern and northeastern sides provides a self-consistent explanation of the observed features of the $\gamma-$ray emission of Puppis A and physically sound values for the CR total energy. 
Trying to explain the northeastern and southwestern SEDs of Puppis A with a single acceleration mechanism (DSA) is much more difficult and problematic (though it cannot be definitely ruled out yet).
Even if Puppis A is a relatively young supernova remnant it is not accelerating particles up to a few TeV, as shown by the non detection with \textit{H.E.S.S.} and the cutoff energy for the particle spectrum at $E < 1$ TeV.

Moreover,  we further delve into the analysis of the $\gamma$-ray source 4FGL J0822.8-4207, located beyond the northern border of Puppis A, and discover a putative new source, namely PS J0824.0-4329, located to the south of the shell. We find out that 4FGL J0822.8-4207 is an extended source and we analyze the SED of the two sources. The analysis shows that both sources are associated with dense interstellar clumps and their $\gamma$-ray emission is the result of a high energy density of cosmic rays therein. Indeed, we show that, considering the diffusion length of CRs escaped from Puppis A, 4FGL J0822.8-4207 and PS J0824.0-4329 can be associated with inelastic proton-proton collisions between hadrons accelerated in Puppis A and the cloud material.

\begin{acknowledgements}
The Fermi LAT Collaboration acknowledges generous ongoing support from a number of agencies
and institutes that have supported both the development and the operation of the LAT as well as scientific data
analysis. These include the National Aeronautics and Space Administration and the Department of Energy in the
United States, the Commissariat à l’Energie Atomique and the Centre National de la Recherche Scientifique / Institut
National de Physique Nucléaire et de Physique des Particules in France, the Agenzia Spaziale Italiana and the Istituto
Nazionale di Fisica Nucleare in Italy, the Ministry of Education, Culture, Sports, Science and Technology (MEXT),
High Energy Accelerator Research Organization (KEK) and Japan Aerospace Exploration Agency (JAXA) in Japan,
and the K. A. Wallenberg Foundation, the Swedish Research Council and the Swedish National Space Board in
Sweden. Additional support for science analysis during the operations phase is gratefully acknowledged from the
Istituto Nazionale di Astrofisica in Italy and the Centre National d’Études Spatiales in France. This work performed
in part under DOE Contract DE-AC02-76SF00515.
We acknowledge Jean Ballet and Fabio Acero for the support they gave on the project. We also acknowledge Juan Luna and Gloria Dubner for providing the \textit{XMM-Newton} maps.
This paper is partially supported by the  Fondazione  ICSC, Spoke 3 Astrophysics and Cosmos Observations. National Recovery and Resilience Plan (Piano Nazionale di Ripresa e Resilienza, PNRR) Project ID CN\_00000013 ``Italian Research Center on  High-Performance Computing, Big Data and Quantum Computing"  funded by MUR Missione 4 Componente 2 Investimento 1.4: Potenziamento strutture di ricerca e creazione di ``campioni nazionali di R\&S (M4C2-19 )" - Next Generation EU (NGEU).
This work was supported by JSPS KAKENHI grant numbers 20KK0309 and 24H00246.
\end{acknowledgements}

\bibliographystyle{aa}
\bibliography{references}

\appendix

\section{SEDs with statistical and systematics error}
We here show SED data points with only statistical (in blue) and only systematics (in red) error bars. The combined plot are shown in Sect. \ref{sec:results}, \ref{sec:discussion} and Fig. \ref{fig:fermi+hess}, \ref{fig:East_West}, \ref{fig:sed_j0822.8}.

\begin{figure}[ht!]
\centering
\includegraphics[width=\columnwidth]{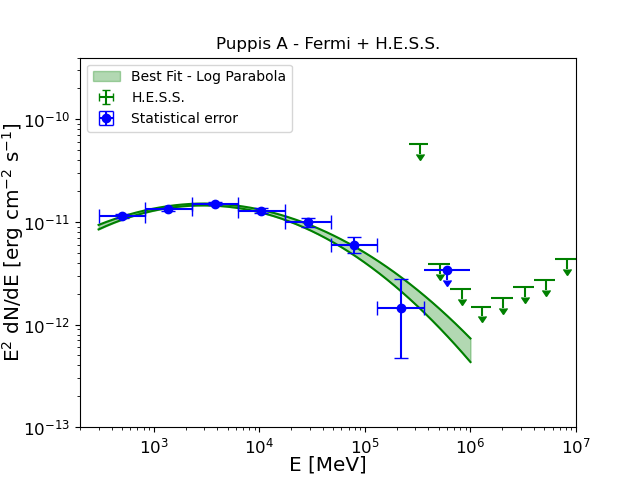}
\includegraphics[width=\columnwidth]{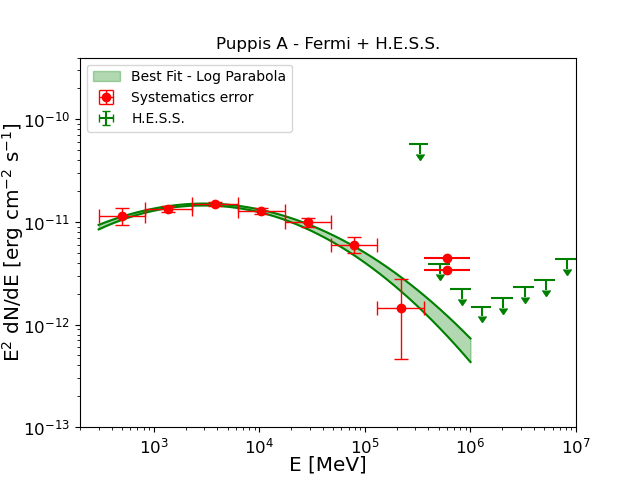}
\caption{SED of Puppis A with the corresponding best fit model and its uncertainties at 68\% confidence level (green curves and shaded region). Fermi upper limits are at 95\% confidence level. The green arrows show the H.E.S.S. upper limits \citep{HessColl15}. \textit{Upper panel} shows statistical errors, \textit{lower panel} shows systematics errors.}
\label{fig:es1}
\end{figure}

\begin{figure}[h!]
\centering
\includegraphics[width=\columnwidth]{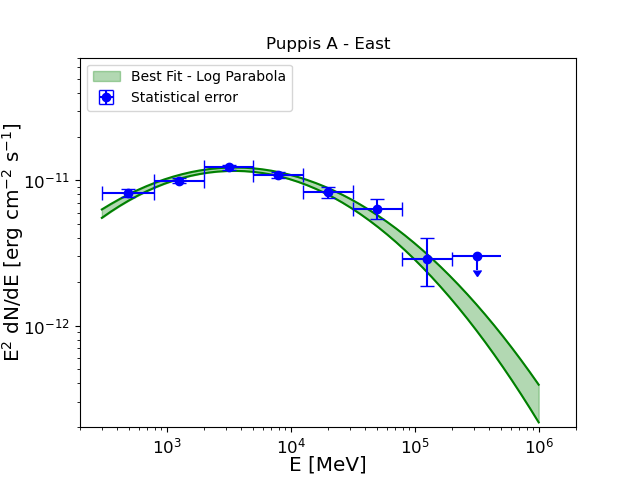}
\includegraphics[width=\columnwidth]{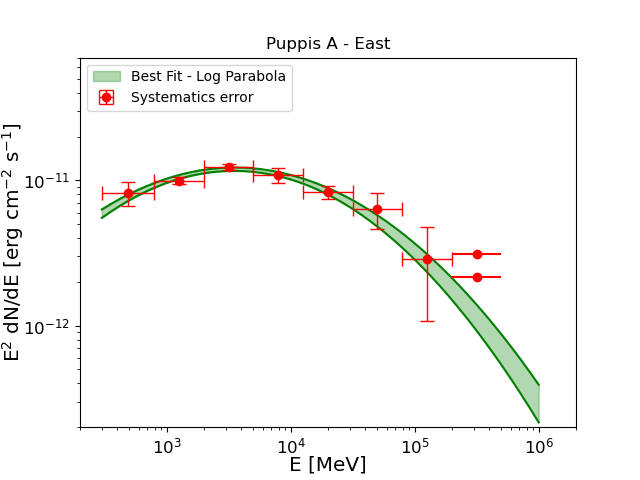}
\caption{SED of the northeastern side of Puppis A, with the corresponding best fit model and its uncertainties at 68\% confidence level (green curves and shaded region). Fermi upper limits are at 95\% confidence level. \textit{Upper panel} shows statistical errors, \textit{lower panel} shows systematics errors.}
\label{fig:es_left}
\end{figure}

\begin{figure}[h!]
\centering
\includegraphics[width=\columnwidth]{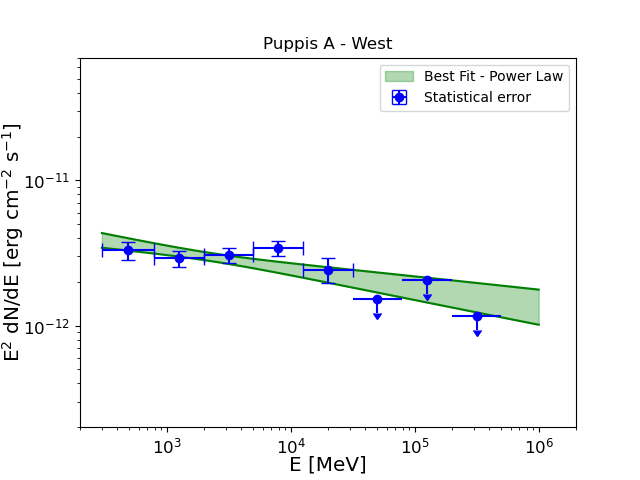}
\includegraphics[width=\columnwidth]{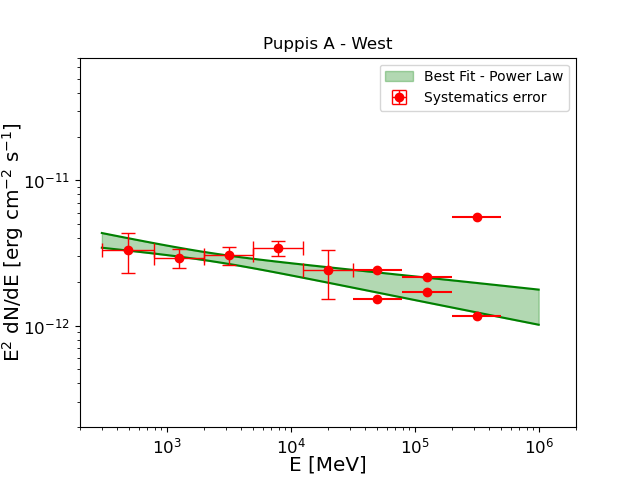}
\caption{SED of the southwestern side of Puppis A, with the corresponding best fit model and its uncertainties at 68\% confidence level (green curves and shaded region). Fermi upper limits are at 95\% confidence level. \textit{Upper panel} shows statistical errors, \textit{lower panel} shows systematics errors.}
\label{fig:es_right}
\end{figure}

\begin{figure}[h!]
\centering
\includegraphics[width=\columnwidth]{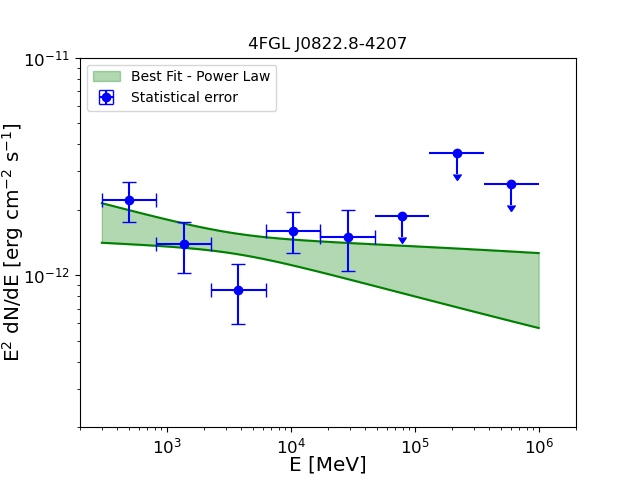}
\includegraphics[width=\columnwidth]{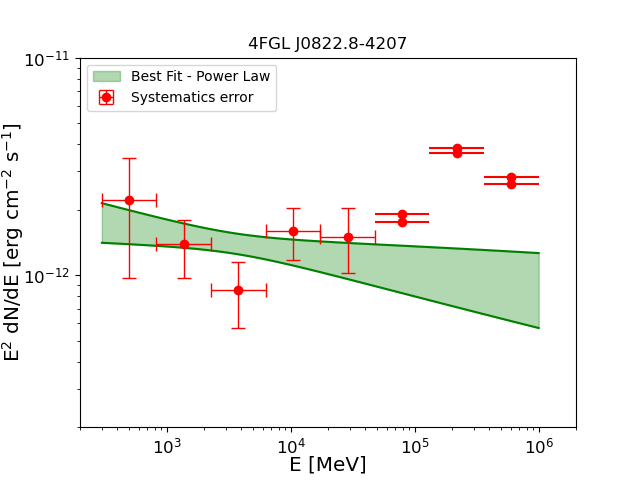}
\caption{SED of the source 4FGL J0822.8-4207 with the corresponding best fit model and its uncertainties at 68\% confidence level (green curves and shaded region). Fermi upper limits are at 95\% confidence level. \textit{Upper panel} shows statistical errors, \textit{lower panel} shows systematics errors.}
\label{fig:araya}
\end{figure}

\begin{figure}[h!]
\centering
\includegraphics[width=\columnwidth]{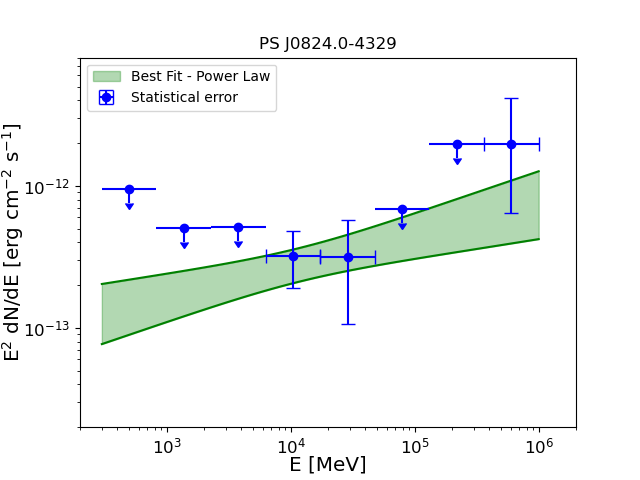}
\includegraphics[width=\columnwidth]{im_app/sed_es1_syst.png}
\caption{SED of the source PS J0824.0-4329. with the corresponding best fit model and its uncertainties at 68\% confidence level (green curves and shaded region). Fermi upper limits are at 95\% confidence level. \textit{Upper panel} shows statistical errors, \textit{lower panel} shows systematics errors.}
\label{fig:ps}
\end{figure}
 
\end{document}